\documentclass{article}

\PassOptionsToPackage{numbers,compress}{natbib}

\usepackage[preprint]{neurips_2025}

\usepackage[utf8]{inputenc}
\usepackage[T1]{fontenc}

\usepackage{amsmath,amssymb,amsfonts}

\usepackage{graphicx}
\usepackage{booktabs}

\usepackage{xcolor}

\usepackage{hyperref}

\usepackage{algorithm}
\usepackage{algorithmic}

\usepackage{listings}

\usepackage{tcolorbox}
\tcbuselibrary{listings,skins,breakable}

\usepackage{enumitem}

\title{FinCARE: Financial Causal Analysis with Reasoning and Evidence}

\author{
  Alejandro Michel \\
  Domyn \\
  New York, US \\
  \texttt{alejandro.zuniga@domyn.com} \\
  \And
  Abhinav Arun \\
  Domyn \\
  New York, US \\
  \texttt{abhinav.arun@domyn.com} \\
  \And
  Bhaskarjit Sarmah \\
  Domyn \\
  Gurgaon, India \\
  \texttt{bhaskarjit.sarmah@domyn.com} \\
  \And
  Stefano Pasquali \\
  Domyn \\
  New York, US \\
  \texttt{stefano.pasquali@domyn.com}
}

\begin{document}
\maketitle
\begin{abstract}
Portfolio managers rely on correlation-based analysis and heuristic methods that fail to capture true causal relationships driving performance. We present a hybrid framework that integrates statistical causal discovery algorithms with domain knowledge from two complementary sources: a financial knowledge graph extracted from SEC 10-K filings and large language model reasoning. Our approach systematically enhances three representative causal discovery paradigms, constraint-based (PC), score-based (GES), and continuous optimization (NOTEARS), by encoding knowledge graph constraints algorithmically and leveraging LLM conceptual reasoning for hypothesis generation. Evaluated on a synthetic financial dataset of 500 firms across 18 variables, our KG+LLM-enhanced methods demonstrate consistent improvements across all three algorithms: PC (F1: 0.622 vs. 0.459 baseline, +36\%), GES (F1: 0.735 vs. 0.367, +100\%), and NOTEARS (F1: 0.759 vs. 0.163, +366\%). The framework enables reliable scenario analysis with mean absolute error of 0.003610 for counterfactual predictions and perfect directional accuracy for intervention effects. It also addresses critical limitations of existing methods by grounding statistical discoveries in financial domain expertise while maintaining empirical validation, providing portfolio managers with the causal foundation necessary for proactive risk management and strategic decision-making in dynamic market environments.
\end{abstract}

\section{Introduction \& Motivation}

Portfolio and risk managers face an increasingly complex challenge: understanding causal mechanisms driving portfolio performance. Current practice relies on heuristic methods and correlation-based analysis; asset managers use experience-based rules, correlation matrices, and traditional factor models that fail to capture causal directionality\cite{rodriguez2025causal, howard2025causal,ma2023linearnonlinearcausalityfinancial,kumar2023causalinferencebankingfinance}. Traditional factor models remain fundamentally descriptive, lacking the scientific grounding necessary to distinguish genuine causal factors from spurious correlations\cite{LopezDePrado2022CausalFactorInvesting,FamaFrench2014FiveFactorModel}. While statistical methods miss theoretically established relationships, knowledge graphs lack empirical validation, and LLMs suffer from the "causal parrot" problem \cite{zecevic2023causal, zhao2024unveiling, jin2024largelanguagemodelsinfer}. Recent empirical studies demonstrate that state-of-the-art LLMs exhibit primarily associative behavior on novel causal questions, lacking the flexible reasoning capabilities required for genuine causal inference \cite{zhao2024unveiling,kıcıman2024causalreasoninglargelanguage}. We present a hybrid framework integrating statistical causal discovery, a financial knowledge graph from SEC filings, and LLM reasoning. Our contributions include: (1) Algorithmic integration of statistical causal discovery with knowledge graph constraints and LLM reasoning while maintaining DAG properties. (2) Substantial improvements across all algorithms: PC (+36\% F1), GES (+100\%), and NOTEARS (+366\%). (3) Reliable counterfactual predictions (MAE: 0.003610) with perfect directional accuracy for interventions, enabling proactive causal analysis for portfolio management.

\section{Related Work}

Causal discovery in finance has evolved with large language models (LLMs). Traditional statistical methods face scalability limitations and the curse of dimensionality.  \cite{li2024causalstock} presented CausalStock, employing functional causal models with lag-dependent temporal discovery for news-driven stock prediction across multiple international markets. \cite{sokolov2023automating} proposed an end-to-end framework leveraging GPT-4 for automated DAG construction, handling hundreds of features through hierarchical clustering. However, this raises the "causal parrot" problem, generating plausible relationships without true understanding \cite{wan2025largelanguagemodelscausal}. \cite{le2025multiagent} addresses this through specialized LLM agents that iteratively refine causal graphs through debate and validation. \cite{ravivanpong2022extracting} validates these theoretical frameworks in practice, showing that neither pure statistical nor pure LLM methods alone suffice for financial causal discovery. The emergence of autonomous causal analysis agents \cite{wang2025causalcopilot, han2024causalagent} further advances the field by providing agents with causal analysis tools and postprocessing modules for DAG tuning, demonstrating how LLMs can serve as decision-making engines for algorithm selection and hyperparameter optimization. Recent hybrid approaches \cite{hiremath2025hybridtopdownglobalcausal} demonstrate that combining constraint-based 
and functional causal modeling can achieve superior accuracy compared to pure methods, 
particularly for sparse causal graphs.

Knowledge graph approaches provide structured causal representations, from hyper-relational graphs \cite{jaimini2022causalkh,10.1145/3397271.3401427} to text extraction methods \cite{anonymous2025ccrag,xu2025fincakg,elhammadi-etal-2020-high}. Recent work has advanced automated KG construction from financial documents; \cite{Li_2024} proposed FinDKG, using fine-tuned LLMs with template-guided extraction to construct dual-structured KGs from financial news. \cite{li2025finkarioeventenhancedautomatedconstruction} extended this with FinKario, demonstrating automated schema construction from equity research reports without predefined domain knowledge. The current frontier fuses LLMs with KGs; \cite{yu2024fusing}'s RC2R framework grounds LLM reasoning in financial KGs through multi-scale contrastive learning, providing rigorous validation via interventional reasoning. Our work leverages FinReflectKG \cite{arun2025finreflectkg}, a KG extracted from SEC 10-K filings using an iterative reflection-driven agentic framework, achieving a 64.8\% compliance score (53.2\% more compared to non-agentic single pass approach). With 227 causal relationships and explicit causal relations (Positively\_Impacts, Negatively\_Impacts, Affects\_Stock), the financial knowledge graph provides domain-grounded priors for our causal discovery pipeline.


\section{Knowledge Graph Integration}
To incorporate domain knowledge into the causal discovery process, we extracted causal relationships from SEC 10-K filings using a structured LLM-based approach. 

This approach is largely inspired by the work done in FinReflectKG, which leverages a comprehensive, business-driven schema. The schema includes 24 categories for entity types (e.g. ORG, PERSON, COMP, etc.) and 27 categories for relationship types (e.g. Has\_Stake\_In, Complies\_With, Invests\_In, etc.). FinReflectKG contains causal relationships as well (e.g. Positively\_Impacts, Negatively\_Impacts, Affects\_Stock, etc.). With FinReflectKG providing useful contextual information about corporate operations and background knowledge, we wanted to further extend these insights to include explicit cause-effect relationships with clear directionality. Hence, we included the development of a specialized causal extraction prompt designed specifically to identify and structure causal relationships from 10-K filings (see Appendix C). This process yielded over 3,000 causal triplets.

\section{Statistical Causal Discovery with Knowledge Graph Constraints}

We evaluate three representative algorithms spanning the major paradigms in causal discovery: constraint-based (PC), score-based (GES), and continuous optimization (NOTEARS). This selection ensures our KG-enhancement approach demonstrates generality across fundamentally different discovery strategies rather than being method-specific. Each algorithm integrates KG constraints through paradigm-appropriate mechanisms.

\subsection{Knowledge Graph Constraint Formulation}

Before describing algorithm-specific integration mechanisms, we formalize how knowledge graph constraints are extracted from 10-K triplets and quantified. Each potential edge $(u,v)$ receives a composite score aggregating three evidence dimensions: strength (confidence labels), frequency (mention count), and coverage (number of companies), which uses the geometric mean:

\begin{equation}
\text{CompositeScore}(u \to v) =(S_{\text{strength}} \cdot S_{\text{freq}} \cdot S_{\text{cov}})^{1/3} 
\end{equation}

where $S_{\text{strength}}$ weights strong and moderate confidence mentions, $S_{\text{freq}}$ normalizes by maximum mentions, and $S_{\text{cov}}$ normalizes by company coverage (see Appendix B for detailed formulations)

Edges are then classified based on composite scores and domain heuristics. High-confidence edges (CompositeScore $>$ 0.4, and mentioned in $>$ 18 companies) become required edges $\mathcal{E}_{\text{req}}$, while low-confidence or rare edges ($<$ 5 mentions) become forbidden edges $\mathcal{E}_{\text{forb}}$. This classification enables a flexible weighting scheme:

\textbf{Edge Weights} $w: \mathcal{V} \times \mathcal{V} \to \mathbb{R}$:

\begin{equation}
w(u,v) =
\begin{cases}
+2.0 & \text{if } (u,v) \in \mathcal{E}_{\text{req}} \\
-2.0 & \text{if } (u,v) \in \mathcal{E}_{\text{forb}} \\
\text{CompositeScore}(u \to v) & \text{if } (u,v) \in \text{KG but neither required nor forbidden} \\
0.0 & \text{otherwise}
\end{cases}
\end{equation}

This encoding allows algorithms to treat KG constraints as soft priors (values in [0,1]) or hard constraints (+/-2.0 for required/forbidden), depending on the algorithmic paradigm.

\subsection{PC Algorithm: Constraint-Based Discovery}

The PC algorithm represents the constraint-based paradigm, discovering causal structure purely through conditional independence testing \cite{10.7551/mitpress/1754.001.0001,10.3389/fgene.2019.00524}. This makes it theoretically sound with minimal parametric assumptions, but also makes it vulnerable to two key failure modes that KG integration addresses: (1) Removing edges with weak statistical signals despite strong domain support (2) Arbitrary edge orientation when independence tests provide insufficient information.

To address these limitations, KG-required edges ($\mathcal{E}_{\text{req}}$ bypass conditional independence testing entirely, while other edges receive adaptive significance thresholds: $\alpha_{\text{adj}}(v_i, v_j) = \alpha \cdot \exp\left(-w(v_i, v_j)\right)$. This makes the algorithm conservative about removing KG-supported edges (e.g., an edge with weight 0.2 requires p < 0.12 for removal, versus an edge with weight 0.8 requires p < 0.067).
Edge orientation uses KG directional preferences when $|w_{ij} - w_{ji}| > 0.2$, resolving ambiguity in constraint-based methods. Based on our sensitivity testing, we use $\alpha$ = 0.15. See Appendix A for algorithm details.

\subsection{GES Algorithm: Score-Based Discovery}

Greedy Equivalence Search (GES) discovers causal structure by optimizing Bayesian Information Criterion (BIC) through local graph modifications, alternating between forward (edge addition) and backward (edge deletion) search phases\cite{10.1162/153244303321897717}. Unlike PC's independence-testing approach, GES is a score-based method that explicitly trades off data fit against model complexity. 

The natural integration point for KG constraints in score-based methods is through scoring function modification. We augment BIC with domain-informed bonuses and penalties that shift the optimization landscape toward theoretically plausible structures without abandoning data fit.

The standard BIC score is modified to incorporate KG structural priors:

\begin{equation}
\text{Score}_{\text{KG}}(G) = \text{BIC}(G) + \lambda_{\text{kg}} \cdot R_{\text{KG}}(G)
\end{equation}

where $R_{\text{KG}}(G)$ is the KG regularization term.

\begin{align}
R_{\text{KG}}(G) &= w_{\text{req}} \cdot \sum_{(u,v) \in \mathcal{E}_{\text{req}}} \mathbb{I}[(u,v) \in G] \cdot w(u,v) \\
&\quad - w_{\text{forb}} \cdot \sum_{(u,v) \in \mathcal{E}_{\text{forb}}} \mathbb{I}[(u,v) \in G] \cdot 10.0
\end{align}

with hyperparameters $\lambda_{\text{kg}} = 10.0$, $w_{\text{req}} = 5.0$, and $w_{\text{forb}} = 20.0$

This means that each KG-required edge included in the graph contributes +50 to the score (10.0 $\times$ 5.0 $\times$ typical w(u,v)), approximately equivalent to improving BIC by 50 points. This makes including domain-supported edges as valuable as finding a substantially better statistical fit. Conversely, each forbidden edge incurs a -200 penalty, effectively prohibiting them unless data evidence is overwhelming.

The modified scoring function acts as a "soft constraint" mechanism because it does not prohibit or require edges (except for forbidden edges), but rather shifts the optimization landscape. If statistical evidence strongly contradicts a required edge, GES can still exclude it, but the algorithm must find a substantially better alternative to overcome the KG bonus. This maintains flexibility while leveraging domain knowledge. See Appendix A for algorithm details.

\subsection{NOTEARS Algorithm: Continuous Optimization}

NOTEARS (Non-combinatorial Optimization via Trace Exponential and Augmented lagRangian for Structure learning) represents a fundamentally different approach: it reformulates discrete DAG structure learning as continuous optimization with a differentiable acyclicity constraint\cite{NEURIPS2018_e347c514}. Rather than searching through discrete graph space like PC and GES, NOTEARS learns a weighted adjacency matrix $W \in \mathbb{R}^{d \times d}$ through gradient-based optimization.

The integration of KG constraints in NOTEARS occurs through three mechanisms operating at different stages of the optimization and post-processing pipeline. Unlike PC's hard constraints or GES's score modification, NOTEARS uses a multi-level approach combining regularization, post-hoc enforcement, and adaptive thresholding. See Appendix A for algorithm details.

\section{Statistical Causal Discovery with LLMs}

While Section 4 detailed how knowledge graph constraints guide causal discovery through extracted 10-K relationships, LLMs offer an alternative source of domain knowledge: reasoning from innate understanding of financial mechanisms. Recent advances in LLM-augmented causal discovery demonstrate the effectiveness of combining statistical methods with LLM reasoning \cite{2024arXiv240501744K,takayama2025integratinglargelanguagemodels}, showing that autonomous systems leveraging both data-driven algorithms and LLM knowledge can outperform conventional approaches on benchmark datasets. The algorithmic integration mechanisms remain identical, the only difference lies in edge provenance.

We implement LLM enhancement using Qwen3-235B-A22B model with thinking mode enabled. Rather than extracting causal relationships from documents, we employ a single reasoning module, MissingEdgeDiscoverer, that generates edge proposals through conceptual reasoning about financial causality (see Appendix C for prompt).

These LLM-generated proposals integrate into PC, GES, and NOTEARS using the exact same mechanisms described in Section 4, with one key substitution: LLM confidence scores replace knowledge graph composite scores as edge weights $w(u,v)$. Both KG and LLM enhancements provide soft prior beliefs about edge existence and direction, differing only in their knowledge source. This unified framework enables clean comparison between document-extracted knowledge (KG), model-internalized knowledge (LLM), and their combination (KG+LLM).

\section{Experiments}

\subsection{Benchmark Dataset}
To benchmark our performance, we developed a realistic synthetic dataset that captures the complex causal relationships present in corporate finance and risk management. The data is created with known causal structures and a ground truth DAG, allowing us to quantitatively measure causal discovery performance through precision, recall, and F1-scores. See Appendix D for data generation details. Our KG demonstrated strong coverage of the ground truth causal structure, capturing 23 of 29 ground truth edges.

\subsection{Causal Graph Recovery}

We evaluate our framework on the synthetic financial dataset of 500 firms across 18 variables with a known ground truth DAG containing 29 edges. We compare four classes of methods: (1) Baseline statistical algorithms without enhancement, (2) KG-Enhanced methods integrating knowledge graph constraints algorithmically, (3) LLM-Only methods using language model reasoning without KG constraints, and (4) KG+LLM methods combining both knowledge graph constraints and LLM reasoning. Each approach is tested across three causal discovery paradigms: constraint-based (PC), score-based (GES), and continuous optimization (NOTEARS).

\begin{figure}[ht]
    \centering
    \makebox[\textwidth][c]{%
        \includegraphics[width=1.0\linewidth]{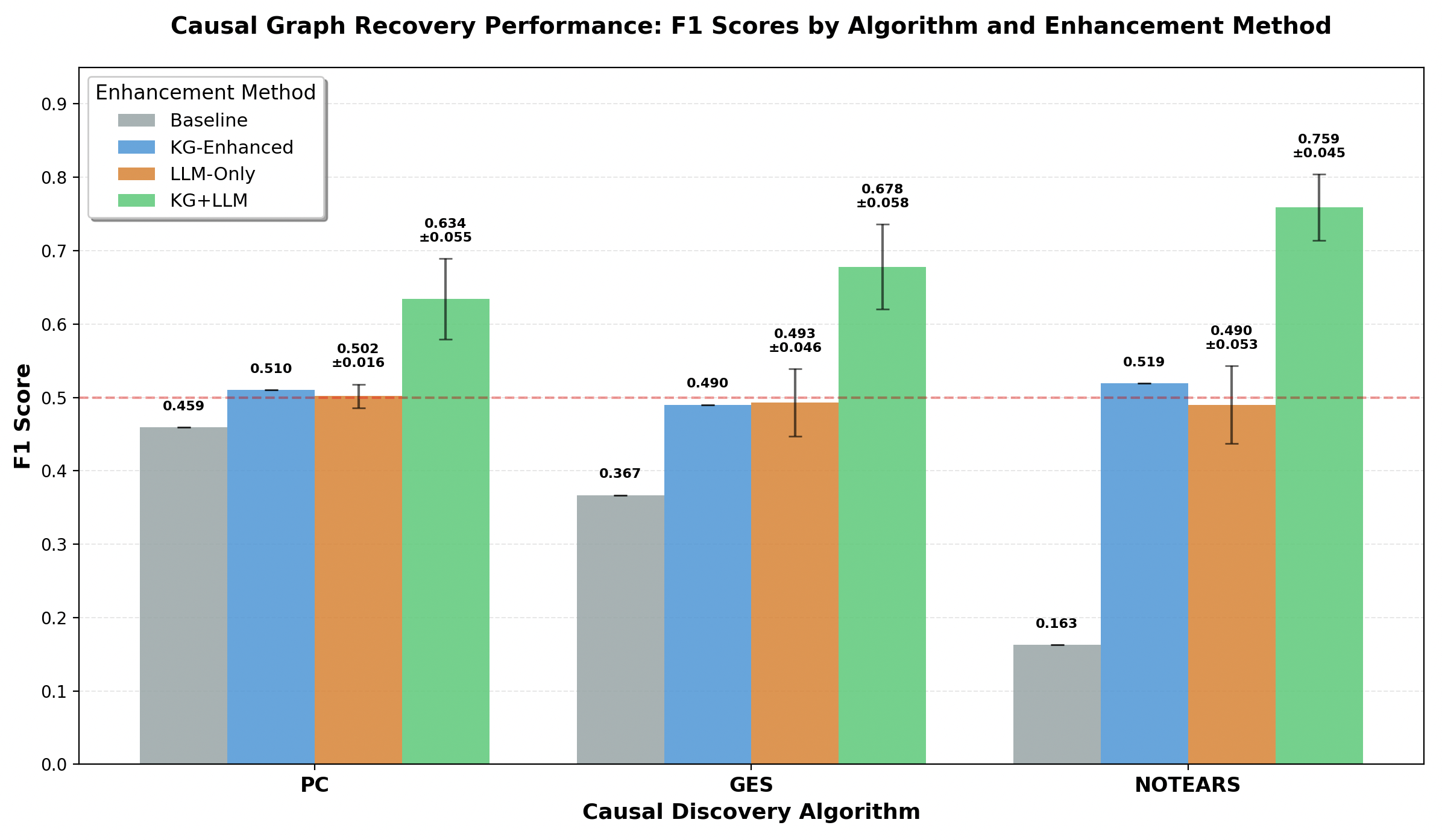}%
    }
    \caption{Graph Recovery Performance Across All Methods ($\pm$1 standard deviation)}
    \label{fig:my_label}
\end{figure}

Table 1 presents comprehensive graph recovery metrics across all methods. To ensure robustness, we report mean performance with standard deviations from 5 independent runs for LLM-enhanced methods (KG+LLM and LLM-Only), while baseline and KG-only methods show deterministic single-run results.

Our results demonstrate clear and consistent performance gains from integrating both knowledge graph constraints and LLM reasoning. In all cases, enhancing any of the three causal algorithms with any of the three strategies (KG-only, LLM-only, or both) led to improvements in graph recovery. In all cases, the best improvement comes when combining KG constraints with LLM reasoning.

\begin{figure}[ht]
    \centering
    \makebox[\textwidth][c]{%
        \includegraphics[width=1.2\linewidth]{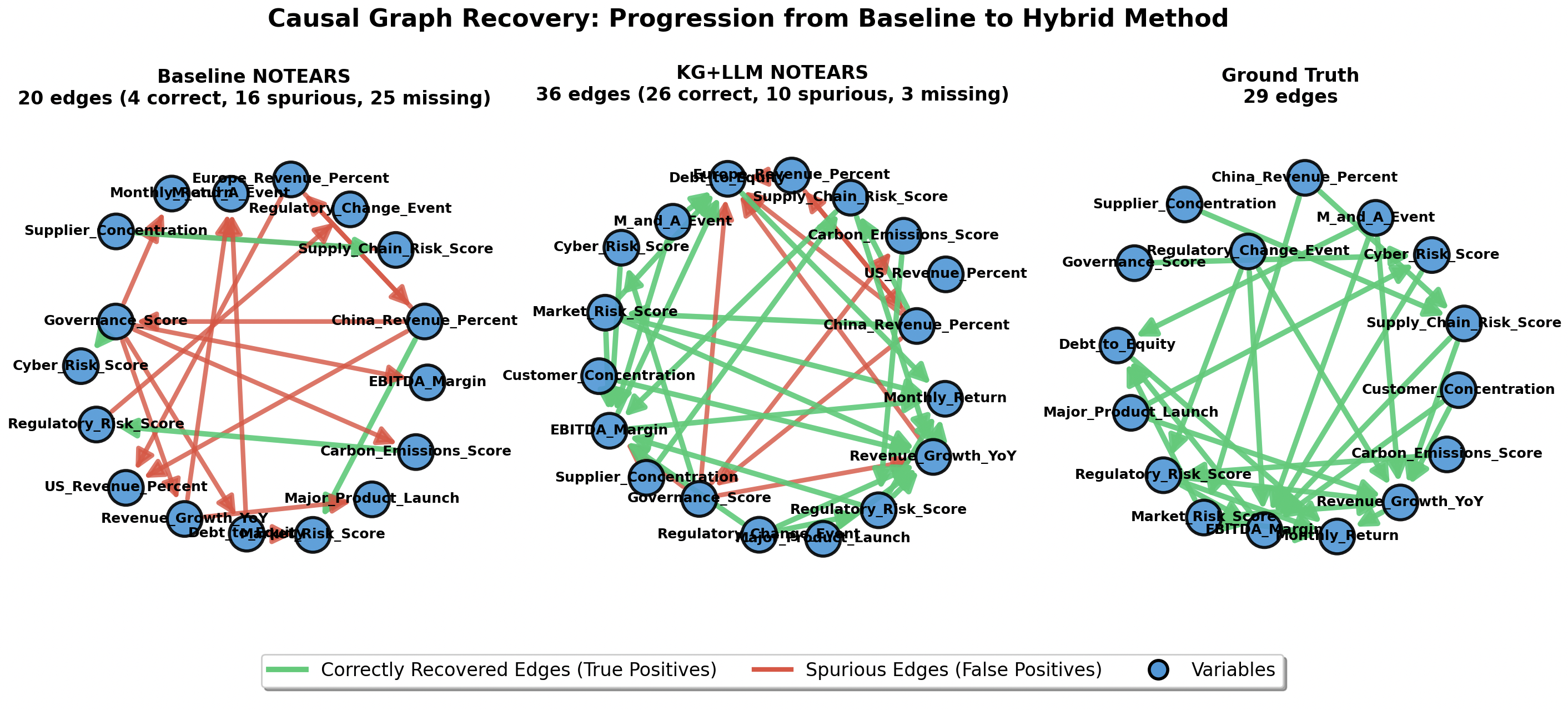}%
    }
    \caption{Causal Graph Recovery Comparison: Baseline vs. KG+LLM vs. Ground Truth}
    \label{fig:my_label}
\end{figure}

Figure 2 illustrates the progression in causal graph recovery from baseline to our hybrid approach. Baseline NOTEARS (left panel) recovers only 4 of 29 ground truth edges (F1=0.163) while KG+LLM-NOTEARS (center panel) achieves near-complete recovery with 26 of 29 edges correctly identified (F1=0.759), capturing the full complexity of financial causal mechanisms.

\subsection{LLM Enhancement Analysis}

So far in our experiments, we have employed the MissingEdgeDiscoverer module as our sole LLM reasoning component. A natural question arises: could we achieve even better performance by expanding our LLM reasoning capabilities? Specifically, we investigate two enhancement strategies: (1) adding additional specialized reasoning modules, and (2) incorporating knowledge graph information directly into LLM prompts. In the following experiments, we only evalaute using NOTEARS.

\subsubsection{Knowledge Graph Prompt Integration}
To rigorously test whether additional LLM reasoning provides value, we design a full system comprising five specialized modules, each targeting different aspects of causal discovery (see Appendix C for prompts):

\begin{itemize}
    \item \textbf{MissingEdgeDiscoverer (already in our system):} Generates targeted hypotheses for 2-3 high-value outcome variables (e.g., \textit{Monthly\_Return}).
    \item \textbf{DataPatternAnalyzer:} Examines distributions, correlations, and sector dependencies to identify patterns statistical methods may overlook.
    \item \textbf{KGRelationshipValidator:} Validates all 11 KG required and 119 forbidden edges by listing them directly in prompts.
    \item \textbf{GraphStructureRefiner:} Proposes specific edge modifications based on graph structural properties.
    \item \textbf{DomainExpertReasoner:} Generates hypotheses from general financial theory without constraint to specific variables.
\end{itemize}

We adopt a leave-one-out ablation approach, systematically disabling each module while keeping others active. To account for LLM output variation, we run each configuration 5 times and report mean performance with standard deviations. Changes smaller than 1$\sigma$ cannot be distinguished from random LLM variation. 

\begin{figure}[ht]
    \centering
    \makebox[\textwidth][c]{%
        \includegraphics[width=1.0\linewidth]{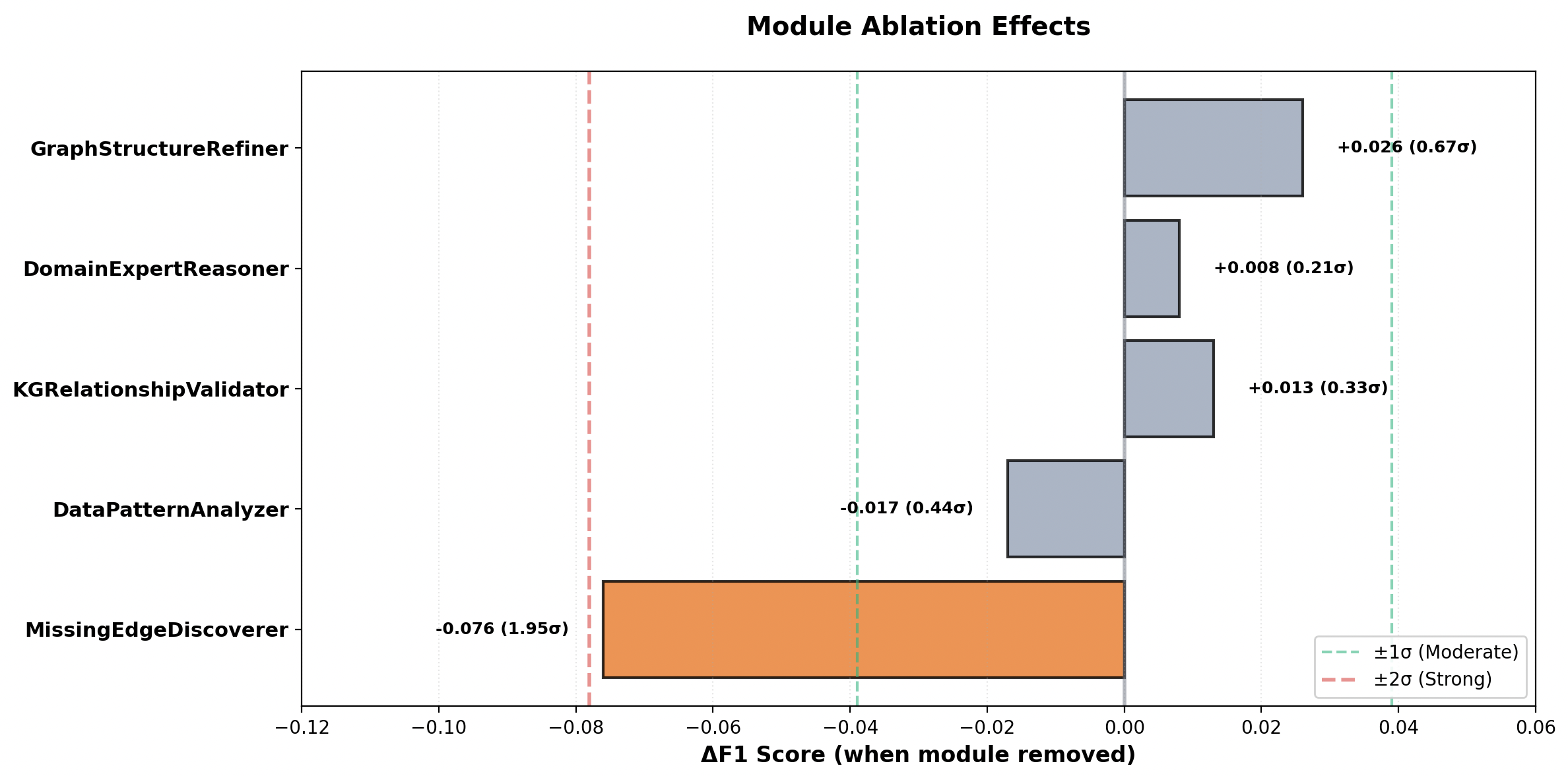}%
    }
    \caption{Leave-One-Out Ablation Analysis of LLM Reasoning Modules on Graph Recovery}
    \label{fig:my_label}
\end{figure}

Figure 3 shows the impact of removing each module. Only MissingEdgeDiscoverer demonstrates statistically meaningful contribution: its removal causes -0.076$\pm$0.018 decrease in F1-score (1.95$\sigma$ deviation). The remaining four modules show changes within natural LLM variation, indicating negligible impact on causal DAG recovery.

The minimal impact of KGRelationshipValidator is particularly instructive. Despite explicitly listing all 11 required and 119 forbidden KG edges in its prompt, removing this module had no significant effect. This suggests that injecting KG information directly into LLM prompts is ineffective compared to encoding KG constraints algorithmically in the optimization objective. 

These findings indicate: (1) a single focused agent outperforms multi-agent approaches for graph recovery, and (2) the optimal strategy combines algorithmic KG constraints with separate LLM-based conceptual reasoning.

\subsection{Counterfactual Estimation}

Counterfactual reasoning in causal models follows the formal framework of structural causal models (SCMs) and Pearl's do-calculus\cite{Pearl_2009, huang2012pearlscalculusinterventioncomplete}, which distinguishes between observational distributions $P(Y \mid X)$ and interventional distributions $P(Y \mid \text{do}(X))$ through graph operations that remove incoming edges to intervention variables. Using the discovered causal structures, we estimate counterfactual outcomes for six financially relevant intervention scenarios (see Appendix E for details). Our counterfactual estimation approach uses linear structural equations fitted to the discovered DAG structure. For each intervention, we set the intervention variable to a specified value and propagate effects through the causal graph following topological order, consistent with the do-operator semantics.

Our KG+LLM-enhanced NOTEARS method achieves an overall mean absolute error (MAE) of 0.003610 when predicting counterfactual outcomes, with perfect directional accuracy (100\%) for all intervention effects

\begin{figure}[ht]
    \centering
    \makebox[\textwidth][c]{%
        \includegraphics[width=1.0\linewidth]{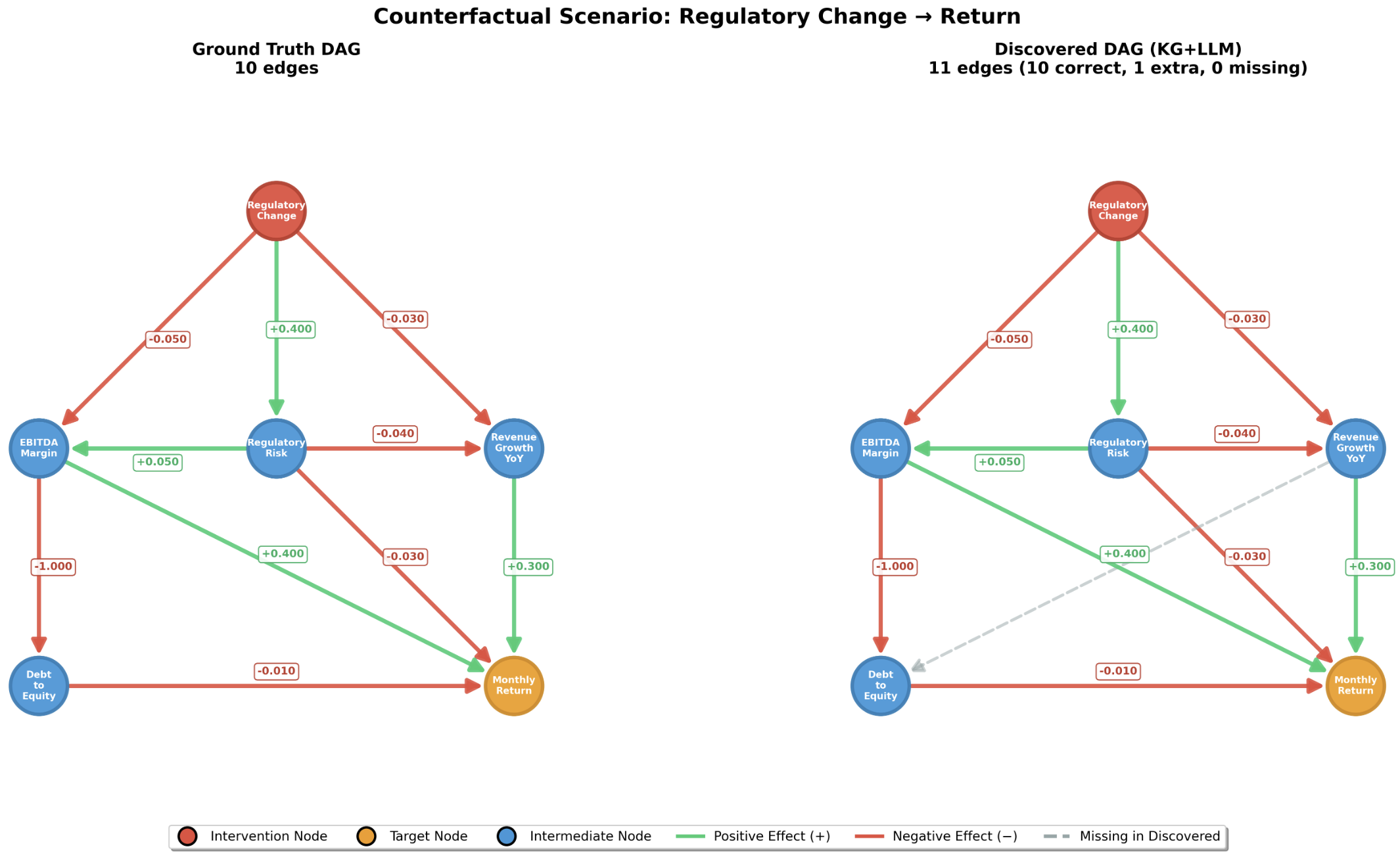}%
    }
    \caption{Counterfactual Estimation for Regulatory Change Intervention}
    \label{fig:my_label}
\end{figure}

To illustrate the framework's practical utility, consider the regulatory change scenario in detail (for more counterfactual examples, see Appendix E). When a major regulatory event occurs (Figure 4, the red node at top), it triggers a +0.4 increase in regulatory risk, meaning the firm's regulatory exposure jumps by 0.4 standard deviations. From here, the effect cascades through three main channels: (1) The Revenue Channel (negative) (2) The Margin Channel (positive) (3) The Direct Sentiment Channel (negative).

Higher regulatory risk depresses revenue growth by -0.040, which then reduces returns by +0.300 per unit of revenue. This creates a net drag of -1.2 basis points on returns through reduced top-line growth. Regulatory risk actually improves EBITDA margins by +0.050 (perhaps through preemptive cost-cutting or operational tightening), which then boosts returns by +0.400 per margin point. This creates a net lift of +2.0 basis points through improved profitability. Finally, regulatory risk directly reduces returns by -0.030 due to investor risk aversion, a -3.0 basis point direct hit. When we sum all pathways, a regulatory change event produces a net -2.2 basis point impact on monthly returns.

\section{Conclusion \& Future Work}
We presented a hybrid framework that systematically integrates three complementary knowledge sources: statistical causal discovery algorithms, domain knowledge extracted from SEC 10-K filings via knowledge graphs, and LLM-driven conceptual reasoning. Our approach demonstrates that all three categories of causal discovery algorithms can greatly benefit from domain knowledge integration, whether it be from knowledge graphs or LLMs, and that the best results come from combining both. The KG+LLM-enhanced NOTEARS method achieved near-complete graph recovery with F1-score of 0.759 compared to baseline performance of 0.163, correctly identifying 26 of 29 ground truth causal edges. Moreover, our framework enables reliable counterfactual reasoning for scenario analysis, with mean absolute error of 0.003610 and perfect directional accuracy for intervention effects. Our ablation studies also revealed critical insights about optimal system design. First, a single, dedicated agent can sometimes be as effective as multiple agents in causal DAG recovery. And second, encoding domain knowledge as algorithmic constraints rather than directly providing knowledge graph information in LLM prompts may provide the best of both knowledge integration methods. Future work could investigate integrating alternative data sources beyond 10-K filings (e.g. earnings calls, news sentiment, supply chain data, etc.), and extending the framework to temporal causal discovery would capture dynamic relationships and lag structures useful for uncovering causal structures in data.

\clearpage

\bibliographystyle{unsrt}
\bibliography{references}

\section*{Appendix A: Algorithmic Knowledge Graph Constraint Details}

\subsubsection*{A1: PC}

\begin{tcolorbox}[title=KG-Enhanced PC Skeleton Discovery, colback=gray!5, colframe=gray!40, enhanced, breakable, left=0pt, top=2pt, bottom=2pt]
\textbf{Input:} Data $X$, variables $V$, significance level $\alpha$, KG-required edges $\mathcal{E}_{\text{req}}$, edge weights $w$.\\
\textbf{Output:} KG-protected undirected skeleton $G_{\text{skel}}$.

\begin{enumerate}[leftmargin=*, label=\arabic*.]
    \item Initialize skeleton with protected edges:
    \begin{itemize}[leftmargin=*, noitemsep]
        \item $\text{protected\_edges} = \{(u,v) : (u,v) \in \mathcal{E}_{\text{req}}\}$
    \end{itemize}
    \item During independence testing phase:
    \begin{itemize}[leftmargin=*, noitemsep]
        \item If $(v_i,v_j) \in \text{protected\_edges}$: \textbf{skip test}
        \item Else, adaptive significance threshold:
        \begin{equation}
        \alpha_{\text{adj}}(v_i, v_j) = \alpha \cdot \exp\left(-w(v_i, v_j)\right)
        \end{equation}
        \item Test $H_0: v_i \perp v_j \mid S$ using partial correlation
        \item Remove edge only if $p$-value $> \alpha_{\text{adj}}$
    \end{itemize}
\end{enumerate}
\end{tcolorbox}

KG-required edges are marked as "protected" and bypass independence testing entirely, and the exponential weighting makes the algorithm more conservative about removing KG-supported edges. For instance, an edge with no KG support receives a weight of 0 and a corresponding alpha of 0.05. But if an edge exists in the KG with a high CompositeScore (and it's not classified as required), then it will recieve an alpha level much lower than 0.05, making the statistical testing much stricter.

This effectively encodes domain knowledge as prior beliefs about edge existence, allowing data to override KG when evidence is sufficiently strong (p<0.018), but protecting theoretically important weak-signal relationships. Edges in $\mathcal{E}_{\text{forb}}$ are removed from the initial complete graph before any testing, reducing the search space and preventing theoretically implausible relationships from entering the skeleton.

\begin{tcolorbox}[colback=gray!5, colframe=gray!40, enhanced, breakable, left=0pt, top=2pt, bottom=2pt]
\textbf{KG-Guided Edge Orientation:}
\begin{enumerate}[leftmargin=*, noitemsep]
    \item For each undirected edge $(v_i,v_j)$ in skeleton:
    \begin{itemize}[leftmargin=*, noitemsep]
        \item Compute directional priors: $w_{ij} = w(v_i,v_j)$, $w_{ji} = w(v_j,v_i)$
        \item Apply threshold $\delta=0.2$:
        \begin{equation}
        \text{Orient as: }
        \begin{cases}
        v_i \to v_j & \text{if } w_{ij} > w_{ji} + \delta \\
        v_j \to v_i & \text{if } w_{ji} > w_{ij} + \delta \\
        \text{Use correlation sign} & \text{otherwise}
        \end{cases}
        \end{equation}
    \end{itemize}
\end{enumerate}
\end{tcolorbox}

\begin{table}[h!]
\centering
\caption{Hyperparameter Sensitivity for PC Algorithm}
\begin{tabular}{ccccc}
\toprule
$\alpha$ value & F1 Score & Precision & Recall & Edges \\
\midrule
0.001 & 0.318 & 0.467 & 0.241 & 15 \\
0.05  & 0.357 & 0.370 & 0.345 & 27 \\
\textbf{0.15} & \textbf{0.459} & 0.438 & 0.483 & 32 \\
0.20  & 0.444 & 0.412 & 0.483 & 34 \\
\bottomrule
\end{tabular}

\vspace{0.5em}
\textbf{Sensitivity Score:} CV(F1) = 0.127
\end{table}

\subsection*{A2: GES}

\begin{tcolorbox}[title=KG-Enhanced GES Edge Scoring, colback=gray!5, colframe=gray!40, enhanced, breakable, left=0pt, top=2pt, bottom=2pt]
\textbf{Input:} DAG $G$, KG constraints $(\mathcal{E}_{\text{req}}, \mathcal{E}_{\text{forb}})$, edge weights $w$, hyperparameters $\lambda_{\text{kg}}, w_{\text{req}}, w_{\text{forb}}$.\\
\textbf{Output:} Score for candidate modifications considering KG priors.

\begin{enumerate}[leftmargin=*, label=\arabic*.]
    \item \textbf{Initialize search with KG seeds:} 
    \begin{itemize}[leftmargin=*, noitemsep]
        \item $G_{\text{init}} = \{(u,v) \in \mathcal{E}_{\text{req}} : w(u,v) > 0.8\}$
    \end{itemize}
    \item \textbf{Forward phase (edge addition):}
    \begin{itemize}[leftmargin=*, noitemsep]
        \item For each candidate edge $(u,v) \notin G$:
        \begin{itemize}[leftmargin=*, noitemsep]
            \item If $(u,v) \in \mathcal{E}_{\text{forb}}$: \textbf{skip}
            \item Else: Evaluate $\text{Score}_{\text{KG}}(G \cup \{(u,v)\})$
            \item If score improvement $> \epsilon$ and acyclic: Add edge
        \end{itemize}
    \end{itemize}
    \item \textbf{Backward phase (edge deletion):}
    \begin{itemize}[leftmargin=*, noitemsep]
        \item Test removing each edge; protected edges incur large penalty
    \end{itemize}
\end{enumerate}
\end{tcolorbox}

\subsection*{A3: NOTEARS}

Four mechanisms work together to address different failure modes: (1) Enforcement prevents weak-signal required edges from being thresholded away (2) Suppression maintains theoretical consistency by blocking implausible relationships (3) Adaptive thresholding provides graduated inclusion criteria based on domain support (4) Cycle resolution ensures DAG constraints while respecting domain knowledge hierarchy.

\begin{tcolorbox}[title=KG-Enhanced NOTEARS Pipeline, colback=gray!5, colframe=gray!40, enhanced, breakable, left=0pt, top=2pt, bottom=2pt]
\textbf{Input:} Learned weight matrix $W$, KG constraints $(\mathcal{E}_{\text{req}}, \mathcal{E}_{\text{forb}})$, edge weights $w$.\\
\textbf{Output:} KG-regularized DAG.

\begin{enumerate}[leftmargin=*, label=\arabic*.]
    \item \textbf{Required Edge Enforcement:} Boost $|W[u,v]|$ for $(u,v) \in \mathcal{E}_{\text{req}}$ below 0.3 to 0.5.
    \item \textbf{Forbidden Edge Suppression:} Set $W[u,v] = 0$ for $(u,v) \in \mathcal{E}_{\text{forb}}$.
    \item \textbf{Adaptive Thresholding:}
    \begin{equation}
    \tau_{ij} = 0.3 \cdot \exp\left(-\frac{w(i,j)}{2}\right)
    \end{equation}
    \item \textbf{Cycle Resolution with KG Priority:} Remove non-required edges in cycles with minimum $|W[u,v]|$; if all required, remove edge with minimum $w(u,v) \cdot |W[u,v]|$.
\end{enumerate}
\end{tcolorbox}

NOTEARS's continuous optimization makes it difficult to integrate hard constraints during learning (unlike discrete search methods). Instead, we apply KG knowledge at post-processing stages where discrete decisions are made, threshold selection, cycle breaking, and final edge inclusion. This preserves NOTEARS's optimization advantages while still leveraging domain expertise.

\begin{table}[h!]
\centering
\caption{Hyperparameter Sensitivity for NOTEARS Algorithm}
\begin{tabular}{ccccc}
\toprule
$\lambda_1$ (L1 penalty) & F1 Score & Precision & Recall & Edges \\
\midrule
0.01  & 0.157 & 0.182 & 0.138 & 22 \\
\textbf{0.05}  & \textbf{0.163} & 0.200 & 0.138 & 20 \\
0.10  & 0.098 & 0.167 & 0.069 & 12 \\
0.50  & 0.054 & 0.125 & 0.034 & 8 \\
\bottomrule
\end{tabular}

\vspace{0.5em}
\textbf{Sensitivity Score:} CV(F1) = 0.461
\end{table}

Table 3 reveals NOTEARS exhibits high sensitivity to the L1 regularization parameter 
$\lambda_1$ (CV = 0.461). 
We use $\lambda_1 = 0.05$ following theoretical guidance.

\section*{Appendix B: Example CompositeScore Calculation}

\begin{equation}
\text{CompositeScore}(u \to v) =(S_{\text{strength}} \cdot S_{\text{freq}} \cdot S_{\text{cov}})^{1/3} 
\end{equation}

where
\begin{align}
S_{\text{strength}} &= \frac{n_{\text{strong}} + 0.5 \cdot n_{\text{moderate}}}{n_{\text{total}}}, \\[2mm]
S_{\text{freq}} &= \frac{n_{\text{mentions}}}{\max(\text{all edge mentions})}, \\[2mm]
S_{\text{cov}} &= \frac{n_{\text{companies}}}{\max(\text{company coverage})}.
\end{align}

Here, $\mathbf{n_{\text{mentions}}}$ represents the total number of times a specific causal relationship appears across all 10-K filings. $\mathbf{n_{\text{strong}}}$ is the number of mentions labeled as "STRONG" confidence by the extraction model. $\mathbf{n_{\text{moderate}}}$ is the number of mentions labeled as "MODERATE" confidence. $\mathbf{n_{\text{total}}}$ is the total mentions for this edge ($\mathbf{n_{\text{strong}}}$ + $\mathbf{n_{\text{moderate}}}$ + $\mathbf{n_{\text{weak}}}$). Finally, $\mathbf{n_{\text{companies}}}$ is the number of distinct companies whose 10-K filings mention this relationship. Example calculations are shown in the appendix.

For example, consider the relationship Market\_Risk\_Score $\rightarrow$  Revenue\_Growth\_YoY extracted from our corpus. This is our most mentioned edge, and has an edge count of 317 total mentions across 91 of 96 companies, with a STRONG, MODERATE, and WEAK strength count of 138, 173, and 6, respectively. The calculations for this edge would be:

\begin{align*}
S_{\text{strength}} &= \frac{138 + 0.5 \cdot 173}{317} = 0.708 && \text{(high-quality evidence)} \\[2mm]
S_{\text{freq}} &= \frac{317}{317} = 1.0 && \text{(most frequent edge in corpus)} \\[2mm]
S_{\text{cov}} &= \frac{91}{96} = 0.948 && \text{(near-universal company coverage)} \\[2mm]
\text{CompositeScore} &= (0.708 \times 1.0 \times 0.948)^{1/3} = 0.876
\end{align*}

This high composite score (0.876 > 0.4 threshold) results in classification as a
required edge, ensuring it receives strong enforcement across all algorithms as detailed below.

Conversely, the edge US\_Revenue\_Percent $\rightarrow$ China\_Revenue\_Percent shows only 27 mentions across 21 companies, yielding S\_{\text{strength}} = 0.48, S\_{\text{frew}} = 0.85, S\_{\text{cov}} = 0.22, and CompositeScore = 0.23. This low-frequency pattern results in classification as a forbidden edge due to likely confounding, preventing algorithms from including this spurious relationship.

\section*{Appendix C: LLM Prompts}

\subsection*{C1: Causal Triplet Extraction Prompt}

Below is the prompt used for the additional extraction of causal triplets from 10-K filings. We applied this extraction methodology to 96 company 10-K filings. For each company, we chunked the 10-K text into manageable segments, then applied the causal extraction prompt to each chunk, and collected all extracted triplets with evidence quotes. During extraction, we made use of Qwen3-235B-A22B model with thinking model enabled. This process yielded a total of 3,419 causal triplets.

\begin{tcolorbox}[
    colback=gray!5!white,
    colframe=gray!75!black,
    rounded corners,
    arc=3mm,
    boxrule=0.5pt,
    breakable,
    fontupper=\small\ttfamily
]
\begin{lstlisting}
You are a financial analyst extracting CAUSAL relationships from SEC 10-K filings. 
IMPORTANT: Only extract relationships where one event/condition CAUSES another. Look for:
- Explicit causal language: "due to", "resulted in", "led to", "caused by", "as a consequence of", "drives", "triggers"
- Conditional causality: "if...then", "when X happens, Y follows"
- Temporal causality: "following X, we experienced Y"

Here is a list of node schemas that nodes could potentially match to (call this TargetColumns):
China_Revenue_Percent,US_Revenue_Percent,Europe_Revenue_Percent,
Governance_Score,M_and_A_Event,Major_Product_Launch,Regulatory_Change_Event,
Supplier_Concentration,Customer_Concentration,Supply_Chain_Risk_Score,
Cyber_Risk_Score,Carbon_Emissions_Score,Regulatory_Risk_Score,
Revenue_Growth_YoY,EBITDA_Margin,Market_Risk_Score,Debt_to_Equity,Monthly_Return

In addition to scanning for relationships, you should also consider converting the natural language and document context to potential entity resolutions. For example, the following would match:
- "Company's ability to attract and retain customers" -> Customer_Concentration
- "higher interest rates" -> Market_Risk_Score
- "global semiconductor industry shortages" -> Supply_Chain_Risk_Score
- "Tariffs" -> Regulatory_Risk_Score
- "weakening of foreign currencies relative to U.S. dollar" -> Market_Risk_Score
- "aggressive price competition" -> Market_Risk_Score
- "EU Digital Markets Act" -> Regulatory_Change_Event
- "demand for the Company's products and services" -> Revenue_Growth_YoY
- "Company's ability to obtain sufficient quantities of components" -> Supplier_Concentration
- "[product] increased in revenue" -> Revenue_Growth_YoY

When it comes to these mappings, you have freedom to think creatively and holistically. 
If something is related to the company, and it could potentially match to one of the TargetColumns, then match it.

For each causal relationship, extract:
{
  "cause_entity": "specific entity/event/condition",
  "cause_entity_potential_column": "one of TargetColumns Value or NONE",
  "cause_type": "METRIC|EVENT|CONDITION|POLICY|MARKET",
  "effect_entity": "specific entity/event/condition", 
  "effect_entity_potential_column": "one of TargetColumns Value or NONE",
  "effect_type": "METRIC|EVENT|CONDITION|POLICY|MARKET",
  "relationship": "one of: [INCREASES, DECREASES, TRIGGERS, PREVENTS, AMPLIFIES, MODERATES]",
  "strength": "STRONG|MODERATE|WEAK",
  "time_lag": "IMMEDIATE|SHORT_TERM|LONG_TERM",
  "evidence": "exact quote from text supporting this causal claim"
}

Text to analyze:
{text_chunk}

Extract all causal relationships as a JSON list:
\end{lstlisting}
\end{tcolorbox}

\subsection*{C2: MissingEdgeDiscoverer}

\begin{tcolorbox}[
    colback=gray!5!white,
    colframe=gray!75!black,
    rounded corners,
    arc=3mm,
    boxrule=0.5pt,
    breakable,
    fontupper=\small\ttfamily
]

\begin{lstlisting}
You are a financial economist with expertise in DIRECT CAUSAL MECHANISMS.

## Target Variable: {target_variable}

## Current Drivers Already Identified:
{sources_to_target if sources_to_target else "None identified"}

## Potential Missing Drivers (with correlations):
{list of variables with correlations}

{pattern_guidance from extracted data patterns}

## CRITICAL INSTRUCTION: Focus on DIRECT Causal Pathways

For {target_variable}, systematically evaluate three types of DIRECT causes:

### TYPE 1: FUNDAMENTAL FINANCIAL DRIVERS (Highest Priority)
If {target_variable} is an outcome variable (Return, Growth, Margin), check:
- **Profitability Metrics** --> Returns/Growth (EBITDA_Margin, Revenue_Growth_YoY)
- **Capital Structure** --> Returns (Debt_to_Equity affects cost of capital)
- **Risk Exposures** --> Returns (Market_Risk, Regulatory_Risk affect discount rates)

QUESTION: Which fundamental financial metrics DIRECTLY determine {target_variable}?
Think: "If this metric improves by 10%, does {target_variable} mechanically change?"

### TYPE 2: RISK TRANSMISSION CHAINS
Risk variables available: {risk_vars}

For risk-to-risk edges, identify DIRECT transmission:
- Geographic exposure --> Market/Supply chain risk
- ESG factors --> Regulatory risk
- Governance quality --> Cyber/Operational risk  
- Concentration --> Specific risk domains

QUESTION: Which operational factors DIRECTLY cause {target_variable} to increase?
Think: "What is the immediate mechanism? No intermediate variables needed?"

### TYPE 3: EVENT-DRIVEN CAUSATION
Events available: {event_vars}

Events (M&A, Product Launch, Regulatory Change) trigger cascades:
- M&A --> Integration risk --> Operational disruption
- Product Launch --> Revenue growth + R&D costs
- Regulatory Change --> Compliance costs --> Margins

QUESTION: Which events would DIRECTLY and immediately affect {target_variable}?
Think: "Does this event have a direct first-order effect?"

## Response Format
Return JSON with ONLY high-confidence, direct causal hypotheses:
{
  "hypotheses": [
    {
      "source": "variable name",
      "target": "{target_variable}",
      "confidence": 0.85,
      "mechanism": "Direct causal mechanism in 1-2 sentences",
      "mechanism_type": "FUNDAMENTAL|RISK_TRANSMISSION|EVENT",
      "expected_coefficient_sign": "POSITIVE|NEGATIVE",
      "expected_strength": "STRONG|MODERATE",
      "alternative_explanations": "What could explain the correlation if not causal?"
    }
  ]
}

**Quality over quantity**: Only propose edges with:
- Confidence >= 0.7
- Clear direct mechanism (no multi-step reasoning chains)
- Strong theoretical or empirical support
\end{lstlisting}
\end{tcolorbox}

\subsection*{C3: DataPatternAnalyzer}

\begin{tcolorbox}[
    colback=gray!5!white,
    colframe=gray!75!black,
    rounded corners,
    arc=3mm,
    boxrule=0.5pt,
    breakable,
    fontupper=\small\ttfamily
]

\begin{lstlisting}
You are an expert econometrician analyzing financial panel data for causal discovery.

## Data Overview
{data_summary - distributions, summary statistics}

## Strong Correlations
{correlations between key variables}

## Variable Categories
- Event Variables: {event_vars}
- Risk Metrics: {risk_vars}  
- Financial Outcomes: {financial_vars}

## Task
Identify potential causal relationships based on:
1. Statistical patterns (correlations, distributions)
2. Temporal logic (events precede outcomes)
3. Financial theory (risk-return relationships)
4. Common confounders that might create spurious correlations

Provide your analysis in JSON format:
{
  "causal_patterns": [
    {
      "source": "variable_name",
      "target": "variable_name",
      "confidence": 0.8,
      "reasoning": "explanation",
      "pattern_type": "temporal/statistical/theoretical"
    }
  ],
  "spurious_correlations": [
    {
      "var1": "variable_name",
      "var2": "variable_name", 
      "likely_confounder": "variable_name or description",
      "reasoning": "why this is likely spurious"
    }
  ],
  "missing_relationships": [
    {
      "description": "relationship that should exist but isn't visible",
      "possible_reason": "why it might be hidden"
    }
  ]
}
\end{lstlisting}
\end{tcolorbox}

\subsection*{C4: KGRelationshipValidator}
\begin{tcolorbox}[
    colback=gray!5!white,
    colframe=gray!75!black,
    rounded corners,
    arc=3mm,
    boxrule=0.5pt,
    breakable,
    fontupper=\small\ttfamily
]

\begin{lstlisting}
You are validating a financial causal graph against domain knowledge.

## Current Graph Edges
{list of current edges in graph}

## KG Required Edges
{list of all 11 required edges from knowledge graph}

## KG Forbidden Edges
{list of all 119 forbidden edges from knowledge graph}

## Statistical Evidence
{correlation and regression statistics for edges}

## Validation Task
For each knowledge graph (KG) relationship, systematically assess:
1. **Causal Plausibility**: Does the direction of the edge make sense theoretically?
2. **Statistical Support**: Do correlations/regressions support this relationship?
3. **Missing Context**: Are there mediating or confounding variables that might explain this edge?
4. **Confidence Level**: How certain are you that this relationship is valid?

Also identify:
- KG edges that should be reversed
- KG edges that should be removed (spurious)
- Important edges missing from the KG

## Response Format
Provide your assessment in JSON format:
{
  "edge_assessments": [
    {
      "source": "variable",
      "target": "variable",
      "kg_status": "required/forbidden",
      "plausibility": 0.8,
      "statistical_support": 0.6,
      "recommendation": "keep/reverse/remove",
      "reasoning": "explanation"
    }
  ],
  "missing_edges": [
    {
      "source": "variable",
      "target": "variable", 
      "confidence": 0.7,
      "reasoning": "why this edge should exist"
    }
  ],
  "overall_kg_quality": 0.7,
  "quality_assessment": "summary of KG quality"
}
\end{lstlisting}
\end{tcolorbox}

\subsection*{C5: GraphStructureRefiner}
\begin{tcolorbox}[
    colback=gray!5!white,
    colframe=gray!75!black,
    rounded corners,
    arc=3mm,
    boxrule=0.5pt,
    breakable,
    fontupper=\small\ttfamily
]

\begin{lstlisting}
You are a causal graph expert refining a financial causal network.

## Current Graph Statistics
- Nodes: {number of nodes}, Edges: {number of edges}
- Density: {graph density}
- Has Cycles: {True/False}
- Components: {number of weakly connected components}

## Potential Issues
- Cycles detected: {number} cycles
- Hub nodes (many outputs): {list of hub nodes}
- Sink nodes (many inputs): {list of sink nodes}

## High-Confidence New Hypotheses
{list of top hypotheses from other modules}

## Refinement Task
Suggest specific modifications to improve causal graph quality:

1. **Cycle Resolution**: Which edges should be removed to eliminate cycles?
2. **Sparsity**: Which weak edges should be removed for a cleaner structure?
3. **Missing Edges**: Which hypotheses should be included?
4. **Direction Corrections**: Which edges need reversal?

Consider:
- Financial causality principles (outcomes don't cause fundamentals)
- Temporal ordering (later events don't cause earlier ones)
- Parsimony (prefer simpler explanations)

## Response Format
Provide recommendations in JSON format:
{
  "edges_to_remove": [
    {
      "source": "variable",
      "target": "variable",
      "reason": "why to remove"
    }
  ],
  "edges_to_add": [
    {
      "source": "variable",
      "target": "variable",
      "confidence": 0.8,
      "reason": "why to add"
    }
  ],
  "edges_to_reverse": [
    {
      "current_source": "variable",
      "current_target": "variable",
      "reason": "why to reverse"
    }
  ],
  "overall_assessment": "summary of graph quality and improvements"
}
\end{lstlisting}
\end{tcolorbox}

\subsection*{C6: DomainExpertReasoner}

\begin{tcolorbox}[
    colback=gray!5!white,
    colframe=gray!75!black,
    rounded corners,
    arc=3mm,
    boxrule=0.5pt,
    breakable,
    fontupper=\small\ttfamily
]

\begin{lstlisting}
You are a senior financial economist with expertise in corporate finance, risk management, and causal inference.

## Current Causal Graph
{list of current edges, limited to top 40}

## Available Variables (by category)
- Events: {event_vars}
- Risk Metrics: {risk_vars}
- Financial Outcomes: {financial_vars}
- Geographic Exposure: {geographic_vars}
- Governance/ESG: {governance_vars}

## Fundamental Financial Theory Frameworks

### 1. Risk-Return Framework
Core principle: {from domain knowledge}
- Risk metrics (market, credit, operational) --> Financial outcomes
- Higher risk exposure --> Required return compensation
- Risk amplification through leverage

### 2. Temporal Causality
Core principle: {from domain knowledge}
- Events and shocks --> Immediate risk adjustments --> Lagged financial impacts
- Corporate actions (M&A, product launches) --> Strategic outcomes
- Regulatory changes --> Compliance costs --> Profitability

### 3. Financial Transmission Mechanisms
Core principle: {from domain knowledge}

**Three Major Transmission Channels:**

a) **Revenue Generation Channel**
   - Market conditions + Customer factors --> Revenue growth
   - Geographic diversification effects on revenue stability
   - Product innovation --> Market share --> Revenue

b) **Profitability Channel**
   - Revenue + Cost structure --> Margins
   - Operational efficiency + Risk management --> EBITDA
   - Concentration risks --> Margin pressure

c) **Return Generation Channel**
   - Fundamental performance (revenue, margins) --> Equity returns
   - Risk exposures --> Return volatility and expectations
   - Financial leverage --> Return amplification/destruction

### 4. Risk Cascade Theory
- **Primary risks** (market, regulatory, cyber) propagate through the system
- **Concentration risks** (customer, supplier) amplify other risks
- **Geographic risks** translate to operational and financial risks

### 5. [Additional frameworks...]

## Task
Based on these general financial principles (NOT specific KG edges), generate broad hypotheses about causal relationships that may exist in the system.

Focus on:
- Theory-driven relationships (not just correlations)
- Mechanisms that explain HOW causation occurs
- Risk propagation pathways
- Financial transmission channels

Provide hypotheses in JSON format:
{
  "domain_hypotheses": [
    {
      "source": "variable",
      "target": "variable",
      "confidence": 0.75,
      "mechanism": "theoretical explanation",
      "framework": "which framework supports this",
      "expected_sign": "POSITIVE/NEGATIVE",
      "strength": "STRONG/MODERATE/WEAK"
    }
  ]
}

Generate 6-10 theoretically grounded hypotheses covering different transmission mechanisms.
\end{lstlisting}
\end{tcolorbox}

\section*{Appendix D: Synthetic Data Generator}

The first step in our data generation process creates panel observations for 500 firms over a span of 60 consecutive quarters (15 years of data). At each quarter, all firms are observed across 18 variables, with common macroeconomic factors (e.g., market conditions, regulatory intensity) varying across time periods while firm-specific characteristics evolve according to theoretically-grounded causal mechanisms. The data generation incorporates realistic causal mechanisms based on established finance theory, including nonlinear relationships, threshold effects, and domain-appropriate noise structures. The last step is then to aggregate this data.

From the generated panel data, we construct a cross-sectional dataset by computing firm-level averages. In total, this yields 500 firm-level observations, each with 18 features. The ground truth causal graph includes 29 edges that represent within-period relationships among these variables. To make the variables directly comparable, we standardize all features using z-score normalization.

This aggregation focuses our analysis on persistent, long-run causal relationships between firm characteristics, with structural associations that differentiate firms from one another. Methodologically, our research question targets cross-sectional structural relationships (e.g. whether firms with certain characteristics such as stronger governance, or higher ESG scores systematically exhibit different risk-return profiles). 

\begin{tcolorbox}[
    colback=gray!5!white,
    colframe=gray!75!black,
    rounded corners,
    arc=3mm,
    boxrule=0.5pt,
    breakable,
    fontupper=\small\ttfamily
]

\begin{lstlisting}
Algorithm: KG-Informed Synthetic Panel Data Generation

Inputs:
- n_firms: number of firms in panel
- n_periods: number of time periods (e.g., months)
- seed: random seed for reproducibility
- include_counterfactuals: whether to generate intervention scenarios

Variables (by group):
- Exogenous: China_Revenue_Percent, US_Revenue_Percent, Europe_Revenue_Percent,
            Governance_Score, Supplier_Concentration, Customer_Concentration,
            Carbon_Emissions_Score
- Event variables: M_and_A_Event, Major_Product_Launch, Regulatory_Change_Event
- Risk metrics: Supply_Chain_Risk_Score, Cyber_Risk_Score, Regulatory_Risk_Score,
                Market_Risk_Score
- Financial outcomes: Revenue_Growth_YoY, EBITDA_Margin, Debt_to_Equity, Monthly_Return

Steps:

1. Initialize generator
   - Set random seed
   - Define all variable groups
   - Prepare storage structures

2. Generate hidden confounders
   - Macro condition: affects all firms over time
   - Industry shocks: firm-specific, time-varying
   - Regulatory intensity: cumulative over time
   - Firm latent quality: firm-specific

3. Generate base panel (per firm, per time period)
   For each firm:
       For each time period t:
           a) Exogenous variables
               - Geographic revenue distribution (normalized to sum 100)
               - Governance, supplier/customer concentration, carbon emissions
           b) Event variables
               - Probabilistic generation influenced by firm quality, governance, regulatory intensity
           c) Risk metrics
               - Supply_Chain_Risk: function of supplier concentration, China exposure, industry shock
               - Cyber_Risk: function of governance score, product launches
               - Regulatory_Risk: function of regulatory events, ESG scores, intensity
               - Market_Risk: function of macro condition, supply chain, China exposure, lagged risk
           d) Financial outcomes
               - Revenue_Growth_YoY, EBITDA_Margin, Debt_to_Equity: influenced by risks, events, lagged values
               - Monthly_Return: ultimate outcome, influenced by all financials, risks, and hidden confounders
           e) Update lagged variables for temporal persistence
   - Add small noise to continuous variables
   - Clip variables to realistic ranges

4. Add temporal dynamics
   - Short-term lags (1-3 periods) for key variables
   - Long-term moving average effects (6-period MA)

5. Generate counterfactual scenarios (optional)
   - Sample firms and periods
   - Apply interventions:
       * Regulatory change
       * Market shock
       * M&A event
       * Supply chain disruption
   - Propagate effects on risks and financials based on KG relationships

6. Return outputs
   - DataFrame of panel data
   - Counterfactual scenarios (if requested)
   - True DAG (list of edges defining the causal structure)


\end{lstlisting}
\end{tcolorbox}

\subsection{Appendix E: Counterfactual Estimation}

\begin{figure}[ht]
    \centering
    \makebox[\textwidth][c]{%
        \includegraphics[width=1.0\linewidth]{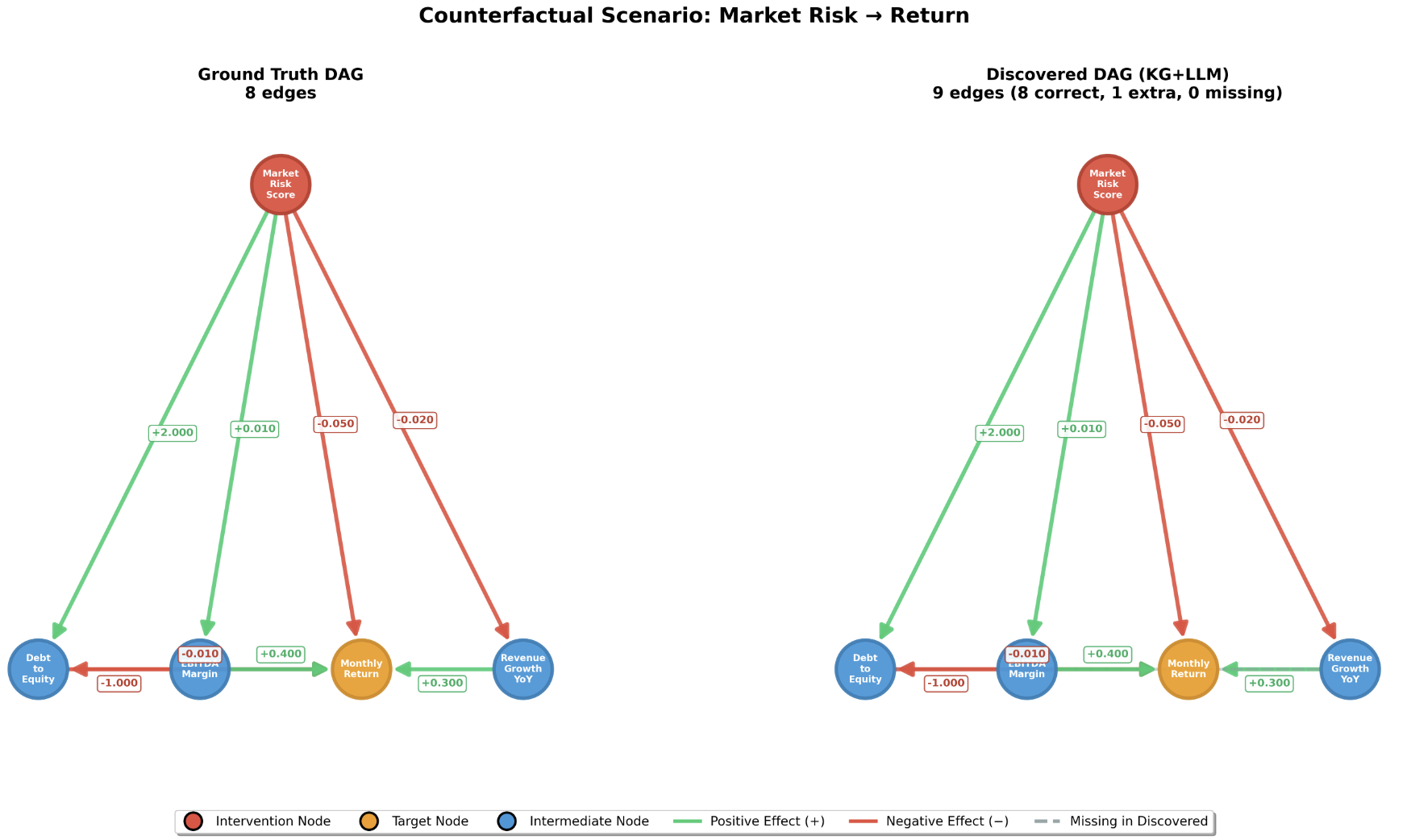}%
    }
    \caption{Counterfactual results on market risk and return}
    \label{fig:my_label}
\end{figure}

\begin{table}[ht]
\centering
\caption{Per-Scenario Performance Summary}
\begin{tabular}{l l l c}
\toprule
\textbf{Scenario} & \textbf{Intervention} & \textbf{Target} & \textbf{MAE} \\
\midrule
Regulatory Change $\rightarrow$ Return & Regulatory\_Change\_Event = 1.0 & Monthly\_Return & 0.000823 \\
M\&A Event $\rightarrow$ Margin & M\_and\_A\_Event = 1.0 & EBITDA\_Margin & 0.005035 \\
Market Risk $\rightarrow$ Return & Market\_Risk\_Score = 0.3 & Monthly\_Return & 0.005869 \\
Supply Chain $\rightarrow$ Revenue & Supply\_Chain\_Risk\_Score = 0.4 & Revenue\_Growth\_YoY & 0.005204 \\
Cyber Risk $\rightarrow$ Margin & Cyber\_Risk\_Score = 0.35 & EBITDA\_Margin & 0.001878 \\
Product Launch $\rightarrow$ Revenue & Major\_Product\_Launch = 1.0 & Revenue\_Growth\_YoY & 0.002854 \\
\midrule
\textbf{Average} & --- & --- & \textbf{0.003610} \\
\bottomrule
\end{tabular}
\label{tab:scenario_performance}
\end{table}

\end{document}